\documentclass[aps,prd,superscriptaddress,eqsecnum,nofootinbib,showkeys,twocolumn,showpacs,preprintnumbers,altaffilletter]{revtex4-1}
\usepackage{graphicx}
\usepackage{euscript,amssymb}
\usepackage{amsfonts}
\usepackage{amsmath}
\usepackage{amssymb}
\usepackage{graphicx}
\usepackage{euscript}
\usepackage[dvips]{color}
\usepackage{hyperref}
\begin{document}
\title{The holographic induced gravity model with a Ricci dark energy:
smoothing the little rip and big rip through Gauss-Bonnet effects?}

\author{Moulay-Hicham Belkacemi}
\email{hicham.belkacemi@gmail.com}
\affiliation{Laboratory of Physics of Matter and Radiation, Mohammed I University, BP 717, Oujda, Morocco}
\author{Mariam Bouhmadi-L\'{o}pez}
\email{mariam.bouhmadi@ist.utl.pt}
\affiliation{Centro Multidisciplinar de Astrof\'{\i}sica - CENTRA, Departamento de F\'{\i}sica, Instituto Superior T\'ecnico, Av. Rovisco Pais 1,1049-001 Lisboa, Portugal}
\author{Ahmed Errahmani}\email{ahmederrahmani1@yahoo.fr}
\affiliation{Laboratory of Physics of Matter and Radiation, Mohammed I University, BP 717, Oujda, Morocco}
\author{Taoufiq Ouali}
\email{ouali\_ta@yahoo.fr}
\affiliation{Laboratory of Physics of Matter and Radiation, Mohammed I University, BP 717, Oujda, Morocco}

\begin{abstract}
We present an holographic brane-world model of the Dvali-Gabadadze-Porrati
(DGP) scenario with and without a Gauss-Bonnet term (GB) in the bulk. We
show that an holographic dark energy component with the Ricci scale as the
infra-red cutoff can describe the late-time acceleration of the universe. In
addition, we show that the dimensionless holographic parameter is very
important in characterising the DGP branches, and in determining the
behaviour of the Ricci dark energy as well as the asymptotic behaviour of
the brane. On the one hand, in the DGP scenario the Ricci dark energy
will exhibit a phantom-like behaviour with no big rip if the holographic
parameter is strictly larger than 1/2. For smaller values, the brane hits a
big rip or a little rip. On the other hand, we have shown that the
introduction of the GB term avoids  the big rip and little rip singularities
on both branches but cannot avoid the appearance of a big freeze singularity
for some values of the holographic parameter on the normal branch, however, these values are very unlikely because they lead to a very negative equation of state at the present and therefore we can speak in practice of singularity avoidance.
At this regard, the equation of state parameter of the Ricci dark energy plays a crucial role, even more important than the GB parameter, in rejecting the parameter space where future singularities appear.
\end{abstract}

\date{\today }
\maketitle



\section{Introduction}

Various astrophysical observations of  type Ia supernovae \cite%
{Perlmutter:1998np}, cosmic microwave background (CMB) \cite{Komatsu:2010fb}
 and large scale structure (LSS) \cite{Tegmark2004}, etc., suggest that the
universe is in accelerating expansion. Quantitative analysis shows that
there is a dark energy (DE) with negative pressure component leading to the
dynamical mechanism of the accelerating expansion of the universe. However,
the nature of this dark energy remains a mystery. Various models of dark
energy have been proposed to solve this problem, such as a small positive
cosmological constant \cite{Peebles} and several kinds of scalar fields such
as quintessence \cite{Ratra}, k-essence \cite{ARMENDARIZ-PICON 2001},
phantom \cite{Caldwell}, etc.

On the other hand, an alternative approach to explain the problem of DE
arises from the holographic principle which states that the number of
degrees of freedom for a system within a finite region should be finite and
bounded by the area of its boundary \cite{'tHooft:1993gx,Susskind:1994vu}.
Furthermore it was also suggested that the ultraviolet (UV) cutoff scale of
a system is connected to its infrared (IR) cutoff scale. Cohen et al. \cite%
{Cohen:1998zx} pointed out that, for a system with size L and UV cutoff,
which is not a black hole, the quantum vacuum energy of the system
should not exceed the mass of the same size black hole, i.e. $L^{3}\rho
_{\Lambda }\leq LM_{P}^{2}$, where $\rho _{\Lambda }$ is the vacuum energy
density caused by UV cutoff $\Lambda $, and $M_{p}$ denotes the Plank mass ($%
M_{p}^{2}=1/8\pi G).$ The largest IR cutoff $L$ is required to saturate this
inequality \cite{Li:2004rb,Hsu:2004ri} \ and its holographic energy density
is given by $\rho _{H}=3c^{2}M_{P}^{2}/L^{2}$, where $c$ is a phenomenological numerical
constant. Later on, we will instead use the parameter $\beta$ defined as $\beta=c^2$. The holographic dark energy model is based on assuming that the
energy density $\rho _{H}$ is responsible for the current speed up of the
universe with $L$ being an appropriate cosmological length. It is well known
that the length that plays the role of the IR cutoff on an holographic
energy density is not unique. One of the choices of the IR cutoff is the
Hubble rate. This choice does not induce acceleration either in a
homogeneous and isotropic universe \cite{Hsu:2004ri} or in the DGP model
\cite{Wu:2007tp}, but it does in the DGP model with a Gauss-Bonnet (GB) term
in the bulk \cite{preparation} (see also Ref.~\cite{saridakis}). In the reference \cite{Li:2004rb}, Li
suggested that by choosing $L$ as the event horizon, the late-time
acceleration of the universe is suitably described. Another choice for the
IR cutoff $L$ was suggested by Gao et al. \cite{Gao} (see also \cite{Nojiri:2005pu}), in which the IR cutoff
of the holographic Ricci dark energy (RDE) was taken to be the Ricci scalar
curvature.

Another approach to explain the observed acceleration of the late
universe is to modify gravity at large scales. Models which describe this
approach can be inspired by the existence of extra dimensions. The simplest
and best studied case is the Dvali-Gabadadze-Porrati (DGP) model \cite%
{Dvali:2000hr}. Such a model contains two kinds of solutions \cite%
{Deffayet:2000uy}: the self-accelerating branch, which suffers from some
problems, and the normal branch. Even though the normal branch is healthy it
cannot describe the current acceleration of the universe unless a dark
energy component is invoked \cite{BouhmadiLopez:2007ts} or the gravitational
action is modified \cite{BouhmadiLopez:2010pp}. Modifying gravity at large
scales by introducing a DGP brane-world model and taking into account
curvature effects, as the GB term, were analysed in {\cite{Kofinas:2003rz},}
{\cite{richard}, \cite{BouhmadiLopez:2008nf} and }\cite{preparation} where
it was shown that they can be quite useful to describe the current
acceleration of the universe. The main aim of this paper is to show if the
holographic RDE is suitable to describe the late time acceleration of the
universe, avoiding at the same time a big rip or little rip in
a DGP model with a Gauss-Bonnet term in the bulk.

The outline of the paper is as follows. In Sect. II we will review briefly
the holographic DGP brane-world model with the Ricci scalar as an IR cutoff.
In Sect. III, we analyse the modified Friedmann equation of the model
without the Gauss-Bonnet term and we find analytically the asymptotical
behaviour of the brane and numerically the whole expansion of the
 brane. In particular, we discuss the behaviour of the equation of
state associated with the holographic Ricci dark energy and we show that for $%
\beta \leq 1/2$, $\beta$ being the holographic parameter, the solutions lead
to a big rip or little rip. This motivate us to consider a
GB term in the bulk action in order to enlarge the $\beta $ spectrum and
overcome this problem. In Sect. IV, we consider the model where the bulk
contains a GB term. We show that the effect of a GB term is to remove the
little rip and big rip singularities on both branches, but it cannot
avoid the appearance of a big freeze singularity for some values of the
holographic parameter $\beta$ on the normal branch, however, these values are very unlikely because they lead to a very negative equation of state at the present and therefore we can speak in practice of singularity avoidance.  At this regard, the equation of state parameter of the Ricci dark energy plays a crucial role, even more important than the GB parameter, in rejecting the parameter space where future singularities appear.
Finally, in Sect. V, we conclude.

\section{ Setup of the model}

\label{sectionII}

We consider a DGP brane-world model, where the bulk action contains a GB
curvature term. The bulk corresponds to two symmetric pieces of a
5-dimensional (5d) Minkowski space-time. The brane is spatially flat and its
action contains an induced gravity term. We assume that the brane is filled
with matter and an holographic dark energy. Then, the modified Friedmann
equation reads \cite{Kofinas:2003rz,richard}:

\begin{equation}
H^{2}=\frac{1}{3M_{P}^{2}}\rho +\frac{\epsilon }{r_{c}}\left( 1+\frac{%
8\alpha }{3}H^{2}\right) H,  \label{friedmann1}
\end{equation}%
where $H$ is the brane Hubble parameter, $\rho =\rho_{\mathrm{m}}+\rho_{%
\mathrm{H}}$ is the total cosmic fluid energy density of the brane which can
be described through a cold dark matter component (CDM) with energy density $%
\rho _{\mathrm{m}}$ and an holographic Ricci dark energy component with
energy density $\rho _{\mathrm{H}}$. The parameters $r_{c}$ and $\alpha$
correspond to the cross over scale and the GB parameter, respectively,
both of them being positive. The parameter $\epsilon$ in Eq.~(\ref{friedmann1})
can take two values: $\epsilon =1$, corresponding to the self-accelerating
branch in the absence of any kind of dark energy \cite{richard}; and $%
\epsilon =-1$, corresponding to the normal branch which requires a dark
energy component to accelerate at late-time \cite%
{preparation,BouhmadiLopez:2010pp,BouhmadiLopez:2008nf}.
For simplicity, we will keep the terminology: (i) self-accelerating branch when $\epsilon=1$ and
(ii) normal branch when $\epsilon=-1$.

As already mentioned in the introduction, $\rho _{\mathrm{H}}$ is related to
the UV cutoff, while $L$ is related to the IR cutoff. We consider a Ricci
dark energy (RDE) \cite{Gao} where the scale $L$ is fixed by $\mathcal{R}/6$%
, $\mathcal{R}$ being the scalar curvature of the brane. For a spatially
flat Friedmann-Lema\^{\i}tre-Robertson-Walker (FLRW) universe
\begin{equation*}
\mathcal{R=}6\left( \dot{H}+2H^{2}\right),
\end{equation*}%
where the dot stands for the derivative with respect to the cosmic time of
the brane. Therefore, the RDE energy density is
\begin{equation}
\rho _{\mathrm{H}}=3\beta M_{P}^{2}\left( \frac{1}{2}\frac{dH^{2}}{dx}%
+2H^{2}\right) ,  \label{HRDE}
\end{equation}%
where $x=-\ln (z+1)=\ln(a)$, $z$ is the redshift and $\beta =c^{2}$ is a
dimensionless parameter that measures the strength of the holographic
component. As we will show, $\beta$ and Eq.~(\ref{friedmann1}) play a
crucial role in determining the expansion of the RDE component. Before
moving forwards, we would like to highlight that we have assumed that the
holographic dark energy stated in Eq.~(\ref{HRDE}) remains valid on
brane-world models like the one we are considering here. A similar approach
was followed in Refs.~\cite{Wu:2007tp,preparation}.

The modified Friedmann equation\ ({\ref{friedmann1}) }can be further
rewritten as:

\begin{equation}
E^{2}=\Omega _{m}e^{-3x}+\beta (\dfrac{1}{2}\dfrac{dE^{2}}{dx}%
+2E^{2})+2\epsilon \sqrt{\Omega _{rc}}(1+\Omega _{\alpha }E^{2})E,
\label{friedmann2}
\end{equation}%
where $E(z)=H/H_{0}$ and
\begin{eqnarray}
\Omega _{m} &=&\frac{\rho _{m_{0}}}{3M_{P}^{2}H_{0}^{2}},\,\,\,\,\Omega
_{r_{c}}=\frac{1}{4r_{c}^{2}H_{0}^{2}},\text{\thinspace and }\,\Omega
_{\alpha }=\frac{8}{3}\alpha H_{0}^{2}\,  \notag \\
&&
\end{eqnarray}%
are the usual convenient dimensionless parameters and the subscripts $0$
denotes the present value (we will follow the same notation as in \cite%
{BouhmadiLopez:2008nf,preparation}). The dimensionless parameter of the
holographic energy density can be written as:

\begin{equation}
\Omega _{\mathrm{H}}=\beta (\dfrac{1}{2}\dfrac{dE^{2}}{dx}+2E^{2}).
\label{OmegaH}
\end{equation}%
From ({\ref{friedmann2})} the variation of the dimensionless Hubble rate $E$
with respect to $x$ is

\begin{equation}
\dfrac{dE}{dx}=-\dfrac{\Omega _{m}e^{-3x}+\left( 2\beta -1\right)
E^{2}+2\epsilon \sqrt{\Omega _{rc}}(1+\Omega _{\alpha }E^{2})E}{\beta E}.
\label{variation of E}
\end{equation}

It is convenient to rewrite the previous equation as
\begin{equation}
\frac{\dot{E}}{H_{0}}=-\frac{\Omega _{m}e^{-3x}+\left( 2\beta -1\right)
E^{2}+2\epsilon \sqrt{\Omega _{rc}}(1+\Omega _{\alpha }E^{2})E}{\beta }.
\label{hdot}
\end{equation}
From this equation, we can conclude that: (i) on the one hand, $\dot{H}$ is
well defined and has no singularity at any finite dimensionless Hubble rate.
This indicates the absence of sudden singularities \cite%
{Shtanov:2002ek,Barrow:2004xh,Nojiri:2005sx,Cattoen:2005dx} in this model. (ii) On the other hand, if $E$
diverges, the energy density $\rho _{\mathrm{H}}$ (cf. Eq.~(\ref{HRDE}))
could also diverge and the brane could undergo a big rip or a big freeze
singularity \cite{bigrip,Nojiri:2005sx,BouhmadiLopez:2006fu} or even a little rip
\cite{Stefancic:2004kb,BouhmadiLopez:2005gk, Frampton:2011sp}. This issue
will become clearer in the next sections.

Even though Eq.~(\ref{hdot}) describes the first derivative of the Hubble
rate, it is not the Raychaudhuri equation but the Friedmann equation (\ref%
{friedmann2}). The appearance of $\dot{H}$ in Eqs.~(\ref{friedmann2}) or (%
\ref{hdot}) is simply a consequence of the definition of the holographic
dark energy (cf. Eq.~(\ref{HRDE})) which depends on the first derivative of
the Hubble rate. The Raychaudhuri equation can be easily obtained from the
modified Friedmann equation (\ref{friedmann2}) and the conservation of the
matter content of the brane. We skip it here as the equation is very
lengthly and not very enlightening.

The modified Friedmann equation (\ref{friedmann2}) evaluated at the present
time implies a constraint on the set of parameters of the model
\begin{equation}
1=\Omega _{m}+\Omega _{\mathrm{H}_{0}}+2\epsilon \sqrt{\Omega _{rc}}%
(1+\Omega _{\alpha }).  \label{constraintfriedmann}
\end{equation}
In addition, we have a couple of constraints on the differential equation (%
\ref{variation of E})
\begin{equation}
\left\{
\begin{array}{c}
E_{0}=E(x=0)=1, \\
dE/dx|_{x=0}=-2+\frac{\Omega _{\mathrm{H}_{0}}}{\beta },%
\end{array}%
\right.  \label{initial conditions}
\end{equation}%
where the second condition comes from the dimensionless energy density
definition (see Eq.~(\ref{OmegaH})) evaluated at the present time. It is
possible to constrain further the cosmological parameter by
using the deceleration parameter, $q=-(1+\frac{1}{E}dE/dx)$, which reads at
present:
\begin{equation}
q_{0}=1-\frac{\Omega _{\mathrm{H}_{0}}}{\beta }.  \label{q0s}
\end{equation}%
The universe is currently accelerating, i.e. $q_{0}<0$, and the holographic
RDE energy density is positive, consequently
\begin{equation}
0<\beta<\Omega _{\mathrm{H}_{0}}.  \label{boundsbeta}
\end{equation}

An analytical solution of Eqs.~(\ref{variation of E}) or (\ref{hdot}) is not
at all obvious, so we will combine an analytical asymptotical analysis with
a numerical one to study the modified Friedmann equation (\ref{variation of E}).
We will also split our analysis in two parts: (i) a purely induced
gravity brane-world model and (ii) the same kind of model with GB
corrections in the bulk.

\section{Holographic Ricci dark energy in a DGP brane-world}

\label{sectionIII}

In order to continue the analysis of this model it is convenient to start
with the simplest case for which \mbox{$\alpha =0$.}

\begin{figure}[t]
\begin{center}
\includegraphics[width=0.7\columnwidth]{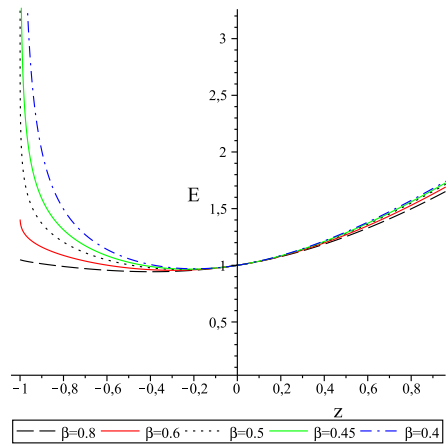}
\end{center}
\caption{Plot of the dimensionless Hubble rate $E$ against the redshift $z$
for the normal branch ($0.43<\beta$) and for the self-accelerating branch ($\beta<0.43$). The cosmological parameters are
assumed to be $\Omega _{m}=0.27$ and $q_{0}=-0.7$. The plot has been done for several values of the holographic parameter $\beta$ that are stated at the bottom of the figure.}
\label{E}
\end{figure}
As we know, the universe is currently accelerating, i.e. $q_{0}<0$. Hence,
using Eqs.~(\ref{constraintfriedmann}) and (\ref{initial conditions}) (for $%
\alpha =0$), we obtain
\begin{equation}
0<\beta =\frac{1-\Omega _{m}-2\epsilon \sqrt{\Omega _{rc}}}{1-q_{0}}<\Omega
_{\mathrm{H}_{0}}.  \label{beta}
\end{equation}%
This inequality is quite enlightening and implies an upper or a lower bound
on the allowed values of $\beta $ depending on which DGP branch we are
considering:
\begin{equation}
\left\{
\begin{array}{c}
\beta <\beta _{\mathrm{lim}}\text{ for \ \ }\epsilon =+1, \\
\beta >\beta _{\mathrm{lim}}\text{ \ for \ }\epsilon =-1,%
\end{array}%
\right.  \label{conditions for beta}
\end{equation}%
where
\begin{equation}
\beta _{\mathrm{lim}}=\frac{1-\Omega _{m}}{1-q_{0}}.  \label{betalim}
\end{equation}%
We can estimate such a bound, $\beta _{\mathrm{lim}}$, taking into account
the latest WMAP7 data \cite{Komatsu:2010fb} and considering that our model
does not deviate too much from the $\Lambda $CDM scenario at the present
time; i.e. $\Omega _{m}\sim 0.27$ and $q_{0}\sim -0.7$, and so we get a
threshold value $\beta _{\mathrm{lim}}\sim 0.43$. From now on, for a given
value of $\beta $, we know with which DGP branch we are dealing with,
depending on whether $\beta $ is smaller or larger than $0.43$.

As we mentioned at the end of Sect.~\ref{sectionII}, it is not at all
obvious how to solve analytically the Friedmann equation (\ref{variation of E}),
even in the simplest case ($\alpha=0$). We will, therefore, analyse the
asymptotical solutions of Eq.~(\ref{variation of E}) and combine them with numerical solutions of the same
equation. In addition, this analysis is very important to analyse the
asymptotical behaviour of the brane and the possible occurrence of any
doomsday in the future.

In the far future $(z\rightarrow -1)$, the energy density of matter $%
\rho_{m}=\rho _{m_{0}}(1+z)^{3}$ is practically zero, and the universe
converges asymptotically to a universe filled exclusively with an
holographic RDE component.

In the absence of matter, we distinguish two cases for the solutions of equation
({\ref{variation of E}) with $\alpha=0$: }
\begin{figure}[t]
\begin{center}
\includegraphics[width=0.7\columnwidth]{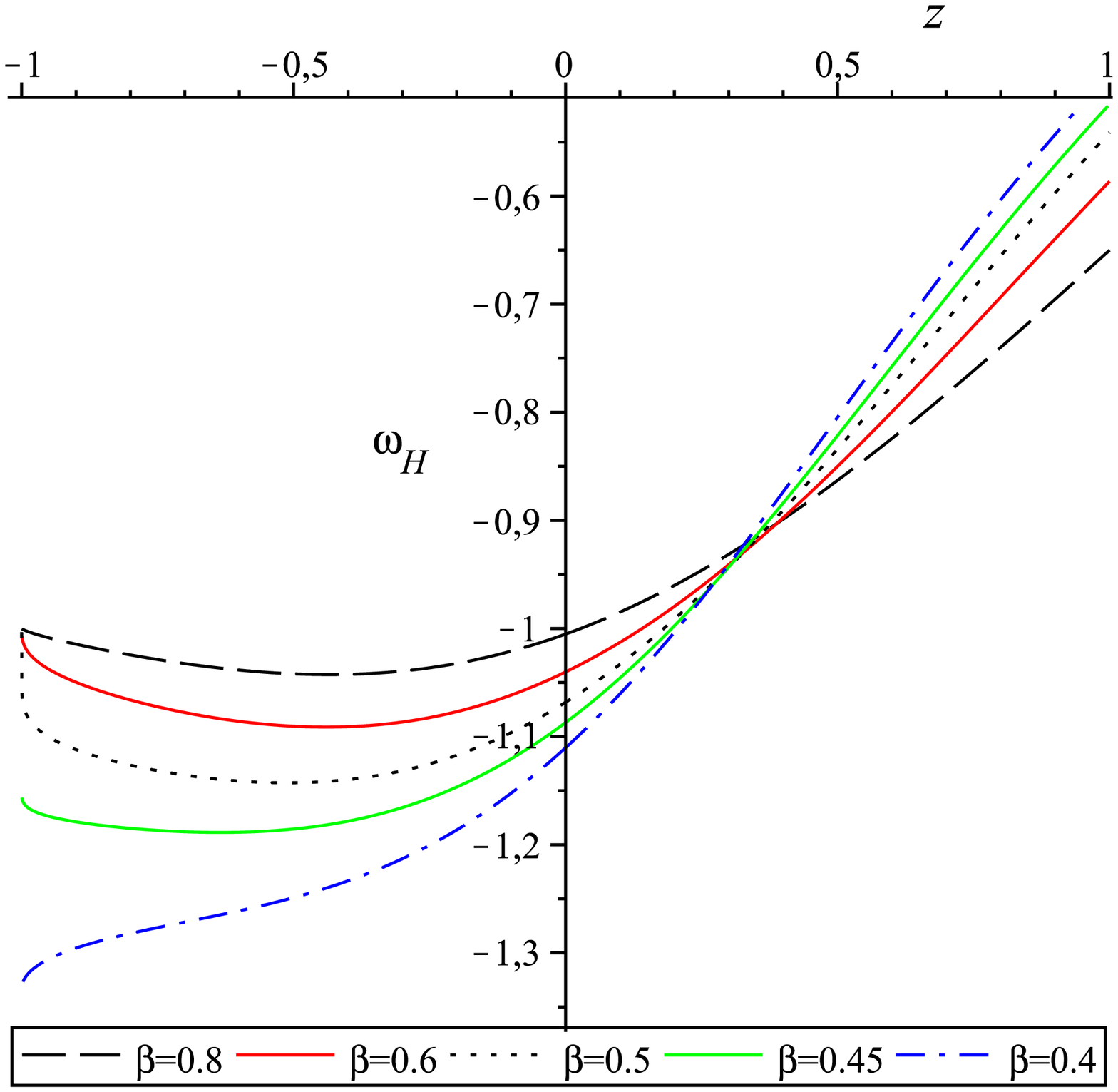}
\end{center}
\caption{Plot of the equation of state parameter $\omega _{_{\mathrm{H}}}$ against the
redshift $z$ for the normal branch ($0.43<\beta$) and for the self-accelerating branch ($\beta<0.43$). The cosmological parameters are assumed to be $\Omega _{m}=0.27$ and $q_{0}=-0.7$. The plot has been done for several values of the holographic parameter $\beta$ that are stated at the bottom of the figure.}
\label{w}
\end{figure}

\subsection{ Asymptotic behaviour \ $\protect\beta =1/2$}

We restrict our analysis to expanding branes. Therefore, there is a unique
solution supported by the normal branch  as implied by the inequalities (\ref{conditions for beta}):
\begin{equation}
E=4\sqrt{\Omega _{rc}}x+ \mathcal{C}_1,  \label{sol beta= 1/2}
\end{equation}%
where $\mathcal{C}_1$ is an integration constant. The solution (\ref{sol
beta= 1/2}) can be further integrated over time
\begin{eqnarray}
x=&\,&\frac{1}{4\sqrt{\Omega_{r_c}}}\left\{(4\sqrt{\Omega_{r_c}}x_1+\mathcal{%
C}_1)\times\right.  \notag \\
&\,&\qquad\quad\left.\exp\left[4\sqrt{\Omega_{r_c}}H_0(t-t_1)\right]-%
\mathcal{C}_1\right\},
\end{eqnarray}
where $x_1$ and $t_1$ are integration constants. Consequently, in the far
future, the scale factor (or $x$) and the Hubble rate diverge. What about $%
\dot H$? It can be easily shown that
\begin{equation}
\frac{dE}{dt}=4\sqrt{\Omega_{r_c}} H,
\end{equation}
and therefore the cosmic derivative of the Hubble rate also diverges. This
kind of event has been named recently a little rip \cite{Frampton:2011sp}
and was previously found on a 4D model \cite{Stefancic:2004kb}
and also on a dilatonic brane-world model \cite{BouhmadiLopez:2005gk}. In
the latter case, the event was called a big rip singularity sent towards an
infinite future cosmic time \cite{BouhmadiLopez:2005gk}.

\begin{figure}[t]
\begin{center}
\includegraphics[width=0.7\columnwidth]{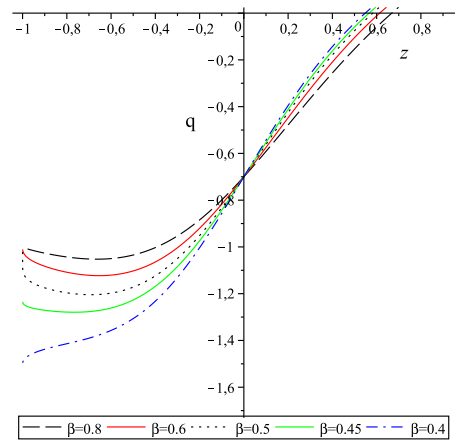}
\end{center}
\caption{Plot of the deceleration parameter $q$ against the redshift $z$ for
the normal branch ($0.43<\beta$) and for the self-accelerating branch ($\beta<0.43$). The cosmological parameters are
assumed to be $\Omega _{m}=0.27$ and $q_{0}=-0.7$. The plot has been done for several values of the holographic parameter $\beta$ that are stated at the bottom of the figure.}
\label{q}
\end{figure}

\subsection{Asymptotic behaviour \ $\protect\beta \neq 1/2$}

The solutions can be expressed as: %
\begin{eqnarray}  \label{sol beta no=1/2}
\arrowvert\frac{2\beta-1}{\beta}E+2\epsilon \frac{\sqrt{\Omega_{r_c}}}{\beta}%
\arrowvert=&\,&|\frac{2\beta-1}{\beta}E_1+2\epsilon \frac{\sqrt{\Omega_{r_c}}%
}{\beta}|  \notag \\
&\,&\exp\left[-\frac{2\beta-1}{\beta}(x-x_1)\right],  \notag \\
\end{eqnarray}%
where $E_1$ and $x_1$ are integration constants, and can be split in two
subsets:

\subsubsection{\ \ \ $1/2<\ \protect\beta $}

This set of values for $\beta$ corresponds to the normal branch as discussed
just after Eq.~(\ref{conditions for beta}), and the dimensionless Hubble
rate takes the solution
\begin{equation}
E_{+}=\frac{2\sqrt{\Omega _{rc}}}{\left( 2\beta-1\right) }  \label{E_{+1}}
\end{equation}
and becomes asymptotically de Sitter in the far future.

\subsubsection{\protect\bigskip\ $\ \ \ \ \protect\beta <1/2$}

This case can correspond to any of the two DGP branches (please see the
argument given just after Eq.~(\ref{conditions for beta})). The
dimensionless Hubble rate blows up in the far future at a finite future
cosmic time, and it follows a super-accelerated expansion until it hits a
big rip singularity \cite{bigrip,Nojiri:2005sx}.

\subsection{Numerical analysis}

Now we return to the equation ({\ref{variation of E}) with $\alpha=0$, in
order to solve it and to study the behaviour of the holographic Ricci dark
energy. As already mentioned, this equation has no analytical solutions, and
a numerical analysis is required. We show our results in Figs.~\ref{E}, \ref%
{w} and \ref{q}. }

In order to solve numerically Eq.~(\ref{variation of E}), we choose $%
\Omega_m\sim 0.27$ and $q_0\sim -0.7$, as given by the latest WMAP7 data
\cite{Komatsu:2010fb} and assuming that our model is very similar to the $%
\Lambda$CDM scenario at the present time. For a given value of $\beta$, the
dimensionless energy density $\Omega _{\mathrm{H}_{0}}$ is fixed through
Eq.~(\ref{q0s}), while the crossover scale parameter $\Omega_{r_c}$ is fixed
by the constraint equation~(\ref{constraintfriedmann}).

In Fig.~\ref{E}, we show five solutions of Eq.~(\ref{variation of E})
corresponding to different values of $\beta$. All the solutions start with a
big bang singularity while their behaviour in the future depends strongly on
the values acquired by $\beta$ (cf. Fig.~\ref{E}): for $\beta =0.4$ and $%
\beta =0.45,$ the brane hits a big rip singularity in the future, for $\beta
=0.5$ the brane expands during an infinite time until it reaches a little
rip, and finally for $\beta =0.6$ and $\beta =0.8$ the brane is
asymptotically de Sitter in the future and the dimensionless Hubble rate $E$
approaches the value $E_{+}=\frac{\beta (1-q_{0})-1+\Omega _{m}}{\left(
2\beta -1\right) }$ (see Eq.~(\ref{E_{+1}})). All these numerical solutions
are in agreement with our analytical analysis as should be.

Even though we have already imposed that the brane is currently
accelerating, more precisely we imposed \mbox{$q_0\sim -0.7$}, it is
important to analyse the behaviour of the equation of state parameter, $%
\omega _{\mathrm{H}}=p_{\mathrm{H}}/\rho _{\mathrm{H}}$, and the
deceleration parameter, $q$, as functions of the redshift. Using the
equation of conservation of the Ricci holographic dark energy (HDE),
\begin{equation}
\dot{\rho}_{\mathrm{H}}+3H(\rho _{\mathrm{H}}+p_{\mathrm{H}})=0,
\label{eq conservation}
\end{equation}%
we obtain
\begin{equation}
\omega _{\mathrm{H}}=-1-\frac13\frac{d\ln\Omega _{\mathrm{H}}}{dx},
\end{equation}%
$\Omega _{\mathrm{H}}$ is defined in equation ({\ref{OmegaH})}. In terms of
the dimensionless Hubble rate $E$ and its derivatives with respect to $x$, $%
\omega _{\mathrm{H}}$ can be rewritten as:
\begin{equation}
\omega _{\mathrm{H}}=-1-\dfrac{\left(\dfrac{dE}{dx}\right)^{2}+E\dfrac{d^{2}E%
}{dx^{2}}+4E\dfrac{dE}{dx}}{3E\dfrac{dE}{dx}+6E^{2}}  \label{eq wH}
\end{equation}
In Fig. \ref{w}, we show an example of the behaviour of the equation of
state parameter, $\omega _{\mathrm{H}}$, for the current cosmological
values, and for different values of $\beta $.

At very late-time $\omega _{\mathrm{H}}$ approaches $-1$ for $\beta \ >1/2$
and tends to $\frac{1}{3}\frac{\beta -2}{\beta }$ for $\beta \ \leq 1/2$ as
can be clearly proved from Eq.~({\ref{sol beta= 1/2}) and Eq.~(\ref{sol beta
no=1/2}).} We would like to highlight in this regard the smooth transition
on the values of $\beta$ when going from an asymptotically de Sitter brane ($%
1/2<\beta$) to a little rip ($\beta=1/2$) to finally a big rip ($\beta<1/2$%
). In addition, in this model we can see the little rip as a division
between an asymptotically  de Sitter universe and a universe
approaching/hitting a big rip singularity.

On the other hand, one can prove that for all\ values of $\beta $ the
equation of state of dark energy will evolve in the future in the region $%
\omega _{\mathrm{H}}\leq-1.$ In addition, using equations ({\ref{variation
of E})} and ({\ref{eq wH})}, the parameter $\omega _{\mathrm{H}}$ can be
expressed at the present time as:

\begin{equation}
\omega _{\mathrm{H}_{0}}=-\frac{\allowbreak \left[ \beta \left(
q_{0}-1\right) -\Omega _{m}\right] (q_{0}-2)-(1+q_{0})\allowbreak }{3\beta
(1-q_{0})}.
\end{equation}

From the current observational data and from (\ref{boundsbeta})
 we can assume that $\beta \leq 0.73,$ which means that $\omega_{\mathrm{H%
}_{0}}<-1,$ and the Ricci dark energy has a phantom-like behaviour. For $%
\beta >0.5,$ this behaviour will continue in the future and will be more and
more like a cosmological constant such that ultimately the universe will
enter  a de Sitter phase in the far future. For $\beta \leq 0.5$ the
behaviour of the Ricci HDE will be phantom-like until the brane reaches a
big rip or a little rip. This behaviour is not consistent with the
holographic principle which prohibits the occurrence of a big rip singularity
\cite{Zhang}. This motivates
us, as  was already done in the case of an holographic dark energy with UV
cutoff $L=H^{-1}$ \cite{preparation}, to introduce a GB term in the bulk to
try to remove this singularity.

Similarly we can plot the deceleration parameter $q$ against the redshift $z$
for different values of $\beta ,$ and for the current cosmological values.
The curves in Fig. \ref{q} show that our universe has entered a phase of
accelerated expansion in the recent past. The deceleration parameter reaches
at very late-time the value $-1$ for $1/2\leq \beta,$ and the value $\frac{%
\beta -1}{\beta }$ for $\beta <1/2.$

\section{Holographic Ricci dark energy in a DGP brane-world with a
Gauss-Bonnet term in the bulk}

We consider now the model where the bulk action contains a Gauss-Bonnet (GB)
curvature term in order to analyse the possibility of avoiding the little
rip and big rip present in the absence of such a term. Our motivation in
including this curvature modification is based on the fact that a GB
curvature term corresponds to an ultra-violet correction to gravity and
therefore can influence the high energy regime where both the little rip and
big rip take place.

By combining the expressions (\ref{constraintfriedmann}) and ({\ref{q0s}),
we obtain:
\begin{equation}
\epsilon\sqrt{\Omega_{r_c}}=\frac{1-\Omega_m-\beta(1-q_0)}{2(1+\Omega_\alpha)%
}.  \label{constraintomrc}
\end{equation}%
Therefore, the right hand side (rhs) of the previous equality is positive
for the self-accelerating branch ($\epsilon=1$) and negative for the normal
branch ($\epsilon=-1$). Consequently, there is a limiting allowed value for $%
\beta$, given by $\beta_{\mathrm{\lim}}$ and defined in Eq.~(\ref{betalim}),
depending on which branch we are considering such that the inequality (\ref%
{conditions for beta}) still holds. Following the same procedure introduced
in Sect.~\ref{sectionIII}, we conclude that $\beta_{\mathrm{\lim}}\sim 0.43$%
. In summary, a value of $\beta$ such that $\beta<0.43$ corresponds to the
self-accelerating branch, while a value of $\beta$ such that $0.43<\beta$
corresponds to the normal branch. }

We would like to highlight that this model has two free parameters: for
example $\Omega_\alpha$ and $\beta$. Indeed, for a given set of values ($%
\Omega_\alpha$,$\beta$), the model is fully defined because $\Omega_m$ is
fixed by observations, $\Omega_{\mathrm{H_0}}$ is determined by Eq.~(\ref%
{q0s}) and $\Omega_{r_c}$ by Eq.~(\ref{constraintomrc}). In our reasoning we
have assumed that the $\Lambda$CDM scenario is a good approximation of our
model at the present time and we have used the latest WMAP7 data \cite%
{Komatsu:2010fb}.

With the inclusion of a GB term in the bulk, it is even less obvious how to
solve analytically the Friedmann equation (\ref{friedmann2}). We will,
therefore, analyse the asymptotical solutions of Eq.~(\ref{friedmann2}) (or
equivalently Eq.~({\ref{variation of E})) and combine them with numerical
solutions of the same equation. This analysis will clarify the possible
avoidance of a doomsday, of the kind of big rip or little rip, present on
the same scenario without GB effects for a set of values of $\beta$. }

In the far future $(z\rightarrow -1)$, the energy density of matter $%
\rho_{m}=\rho _{m_{0}}(1+z)^{3}$ can be neglected, and the universe
converges asymptotically to a universe filled exclusively with an
holographic RDE component.

\begin{figure}[t]
\begin{center}
\includegraphics[width=0.7\columnwidth]{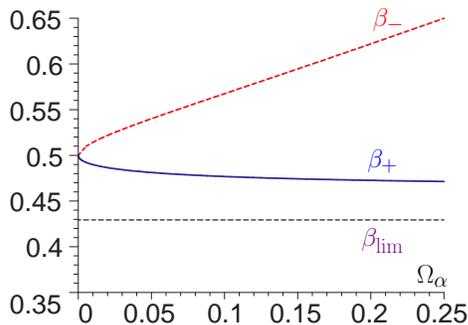}
\end{center}
\caption{Plot of the parameters $\protect\beta_\pm$ and $\protect\beta_{%
\mathrm{lim}}$, defined in Eqs.~(\protect\ref{defbetapm}) and (\protect\ref%
{betalim}), respectively, versus the parameter $\Omega_\protect\alpha$. We
have used the current observational data $q_{0}\sim -0.7,$ and $\Omega
_{m}\sim 0.27$. The parameter $\protect\beta_{\mathrm{lim}}$ defines the
border line between the normal branch ($\protect\beta_{\mathrm{lim}}<\protect%
\beta$) and the self-accelerating branch ($\protect\beta<\protect\beta_{%
\mathrm{lim}}$). For a given value of $\protect\beta$, the parameters $%
\protect\beta_\pm$ fix the sign of the discriminant $\mathcal{D}$ (cf.
Eq.~(\protect\ref{factorD})).}
\label{plotbetas}
\end{figure}

\subsection{Asymptotic behaviour}

In the absence of matter, the modified Friedmann equation ({\ref{variation of E}%
) can be rewritten as:
\begin{equation}
\beta\dfrac{dE}{dx}=-\left[{\left( 2\beta -1\right) E+2\epsilon \sqrt{\Omega
_{rc}}(1+\Omega _{\alpha }E^{2})}\right],  \label{FriedasympGB}
\end{equation}
whose solution depends on the sign of the discriminant $\mathcal{D}$ of the
polynomial on the rhs, which reads
\begin{eqnarray}
\mathcal{D}=\left( 2\beta -1\right) ^{2}-16\Omega _{r_c}\Omega _{\alpha }.
\label{defD}
\end{eqnarray}
It is convenient to factorise the discriminant $\mathcal{D}$ as follows
\begin{equation}
\mathcal{D}=\mathcal{F}(\beta-\beta_-)(\beta-\beta_+).  \label{factorD}
\end{equation}
where
\begin{eqnarray}
\beta_\pm&=&\frac{1+\Omega_\alpha\pm2\sqrt{\Omega_\alpha}(1-\Omega_m)}{2%
\left[1+\Omega_\alpha\pm\sqrt{\Omega_\alpha}(1-q_0)\right]},
\label{defbetapm} \\
\mathcal{F}&=&4\left[1-\Omega_\alpha\left(\frac{1-q_0}{1+\Omega_\alpha}%
\right)^2\right].  \label{coeffF}
\end{eqnarray}
We show in Fig.~\ref{plotbetas}, the shape of $\beta_\pm$ for physically
reasonable values of $\Omega_\alpha$. It can be shown as well  that for the
same range of values of $\Omega_\alpha$, the coefficient $\mathcal{F}$ is
positive (see Eq.~(\ref{coeffF})). Therefore, the discriminant $\mathcal{D}$
is positive when $\beta<\beta_+$ or $\beta_-<\beta$, negative for $%
\beta_+<\beta<\beta_-$ and vanishes for $\beta=\beta_+,\beta_-$. We next
tackle each of these cases separately. }

\subsubsection{Negative discriminant ($\mathcal{D}<0$)}

The discriminant $\mathcal{D}$ is negative when $\beta$ satisfies $%
\beta_+<\beta<\beta_-$ (cf. Eq.~(\ref{factorD}) and Fig.~\ref{plotbetas}).
As $\beta_{\mathrm{lim}}<\beta_+$ this case can take place only for the
normal branch and the Friedmann equation (\ref{FriedasympGB}) can be
integrated as:

\begin{equation}
E=\frac{1}{4\Omega _{\alpha }\sqrt{\Omega _{rc}}}\left\{ \sqrt{\left\vert
\mathcal{D}\right\vert }\tan \left[ \frac{\sqrt{\left\vert \mathcal{D}%
\right\vert }}{2\beta }\left( x-\mathcal{C}_{2}\right) \right] +2\beta
-1\right\} ,  \label{solution D<0}
\end{equation}%
where
\begin{equation}
\mathcal{C}_{2}=x_{2}+\frac{2\beta }{\sqrt{\left\vert \mathcal{D}\right\vert
}}\arctan \left[ \frac{-4\Omega _{\alpha }\sqrt{\Omega _{r_{c}}}E_{2}+2\beta
-1}{\sqrt{\left\vert \mathcal{D}\right\vert }}\right] ,
\end{equation}%
and $x_{2}$ and $E_{2}$ are integration constants. Consequently, there is
always a finite value of the scale factor or $x$ where the Hubble rate blows
up:
\begin{equation}
x_{\mathrm{sing_{1}}}=\mathcal{C}_{2}+\frac{2\beta }{\sqrt{\left\vert
\mathcal{D}\right\vert }}(n+\frac{1}{2})\pi ,\quad n\in \mathbb{Z}.
\end{equation}%
The singularity takes place at $x_{\mathrm{sing}}$ for an integer $n$ such
that $x_{\mathrm{sing_{1}}}$ takes the smallest possible positive value. Not
only the Hubble rate but also its cosmic derivative blow up at $x_{\mathrm{%
sing_{1}}}$ (see Eq.~(\ref{hdot})). Finally, integrating the dimensionless
Hubble rate (\ref{solution D<0}) with respect to time we obtain
\begin{eqnarray}
&\,&\sqrt{\left\vert \mathcal{D}\right\vert }\ln \left\{ \frac{\sqrt{%
\left\vert \mathcal{D}\right\vert }\tan \left[ \frac{\sqrt{\left\vert
\mathcal{D}\right\vert }}{2\beta }\left( x-\mathcal{C}_{2}\right) \right]
+2\beta -1}{\sqrt{1+\tan ^{2}\left[ \frac{\sqrt{\left\vert \mathcal{D}%
\right\vert }}{2\beta }\left( x-\mathcal{C}_{2}\right) \right] }}\right\}
+2\beta -1  \notag \\
&&\quad =\frac{H_{0}\left[ \left\vert \mathcal{D}\right\vert +(2\beta -1)^{2}%
\right] }{4\Omega _{\alpha }\sqrt{\Omega _{r_{c}}}}(t-t_{\mathrm{sing_{1}}}),
\end{eqnarray}%
concluding therefore that the brane hits a big freeze singularity in the
future as the event $x_{\mathrm{sing_{1}}}$ takes place at a finite future
cosmic time $t_{\mathrm{sing_{1}}}$ \cite{Nojiri:2005sx,BouhmadiLopez:2006fu}%
.


\begin{figure}[t]
\begin{center}
\includegraphics[width=0.7\columnwidth]{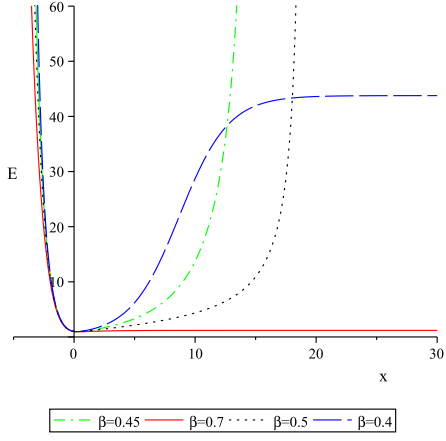}
\end{center}
\caption{Plot of the dimensionless Hubble rate $E$ against $x$ for the normal branch ($0.43<\beta$) and for the self-accelerating branch ($\beta<0.43$). The cosmological parameters are assumed to be $\Omega _{m}=0.27$ and $q_{0}=-0.7$. The plot has been done for several values of the holographic parameter $\beta$ that are stated at the bottom of the figure. The reason we show this plot in terms of $x$, rather than $z$, is to see clearly the presence of a big freeze singularity for some values of $\beta$.}
\label{E with GB}
\end{figure}


\subsubsection{Positive discriminant ($0<\mathcal{D}$)}

The discriminant $\mathcal{D}$ is positive when $\beta<\beta_+$ or $%
\beta_-<\beta$ (cf. Eq.~(\ref{factorD}) and Fig.~\ref{plotbetas}) which can
take place either on the normal branch ($\beta>\beta_{\mathrm{%
lim}}$) or the self-accelerating branch ($\beta_{\mathrm{lim}}>\beta$).

The solution of the asymptotic Friedmann equation (\ref{FriedasympGB})
reads:
\begin{equation}
\left\vert\frac{4\epsilon \Omega_\alpha\sqrt{\Omega_{r_c}}\, E+2\beta-1-%
\sqrt{\mathcal{D}}} {4\epsilon \Omega_\alpha\sqrt{\Omega_{r_c}}\, E+2\beta-1+%
\sqrt{\mathcal{D}}} \right\vert= \mathcal{C}_2\exp\left[-\frac{\sqrt{%
\mathcal{D}}}{\beta}(x-x_2)\right],  \label{solution D>0}
\end{equation}
where
\begin{equation}
\mathcal{C}_2=\left\vert\frac{4\epsilon \Omega_\alpha\sqrt{\Omega_{r_c}}\,
E_2+2\beta-1-\sqrt{\mathcal{D}}} {4\epsilon \Omega_\alpha\sqrt{\Omega_{r_c}}%
\, E_2+2\beta-1+\sqrt{\mathcal{D}}} \right\vert,
\end{equation}
$x_2$ and $E_2$ are integration constants. We can then deduce
that the brane expands till it reaches an infinite size; i.e. for very large
values of $x$, if and only if the brane is asymptotically de Sitter with Hubble rate
\begin{equation}
E_+=\frac{\sqrt{\mathcal{D}}-(2\beta-1)}{4\epsilon\Omega_\alpha\sqrt{%
\Omega_{r_c}}}.  \label{E_{+2}}
\end{equation}
Notice that ${E_{+}}$ reduces to Eq.~(\ref{E_{+1}}) in absence of GB effects
in the bulk. The brane can be asymptotically de Sitter only in too cases:
(i) For $\epsilon=1$ as $\beta<\beta_{\mathrm{lim}}<\frac12$ and (ii) for $%
\epsilon=-1$ if $\frac12<\beta_-< \beta$. In both cases the Hubble
rate $E_+$ is a positive constant.

What happens in the remaining case; i.e. $\epsilon=-1$ and $\beta_{\mathrm{%
lim}}<\beta<\beta_+$? In the latter case $E_+$ is negative and the asymptotic
analysis performed is no longer valid. In fact, the brane faces a big freeze
singularity in this situation where the expansion of the brane can be
approximated by
\begin{equation}
E\sim\frac{1}{2\Omega_{\alpha}\sqrt{\Omega_{r_c}}\,(x_{\mathrm{sing_2}}-x)}.
\label{solution2 D>0}
\end{equation}
The constant $x_{\mathrm{sing_2}}$ stands for the ``size'' of the brane at
the big freeze singularity. The Hubble rate (\ref{solution2 D>0}) can be
integrated over time, resulting in the following expansion for the scale
factor of the brane
\begin{equation}
a=a_{\mathrm{sing_2}}\exp\left[-\left(\frac{H_0}{2\Omega_{\alpha}\sqrt{%
\Omega_{r_c}}}\right)^{\frac12}\sqrt{t_{\mathrm{sing_2}}-t}\right]
\end{equation}
The big freeze singularity takes place at a finite scale factor, $a_{\mathrm{%
sing_2}}$, and a finite cosmic time, $t_{\mathrm{sing_2}}$.


\begin{figure}[t]
\begin{center}
\includegraphics[width=0.7\columnwidth]{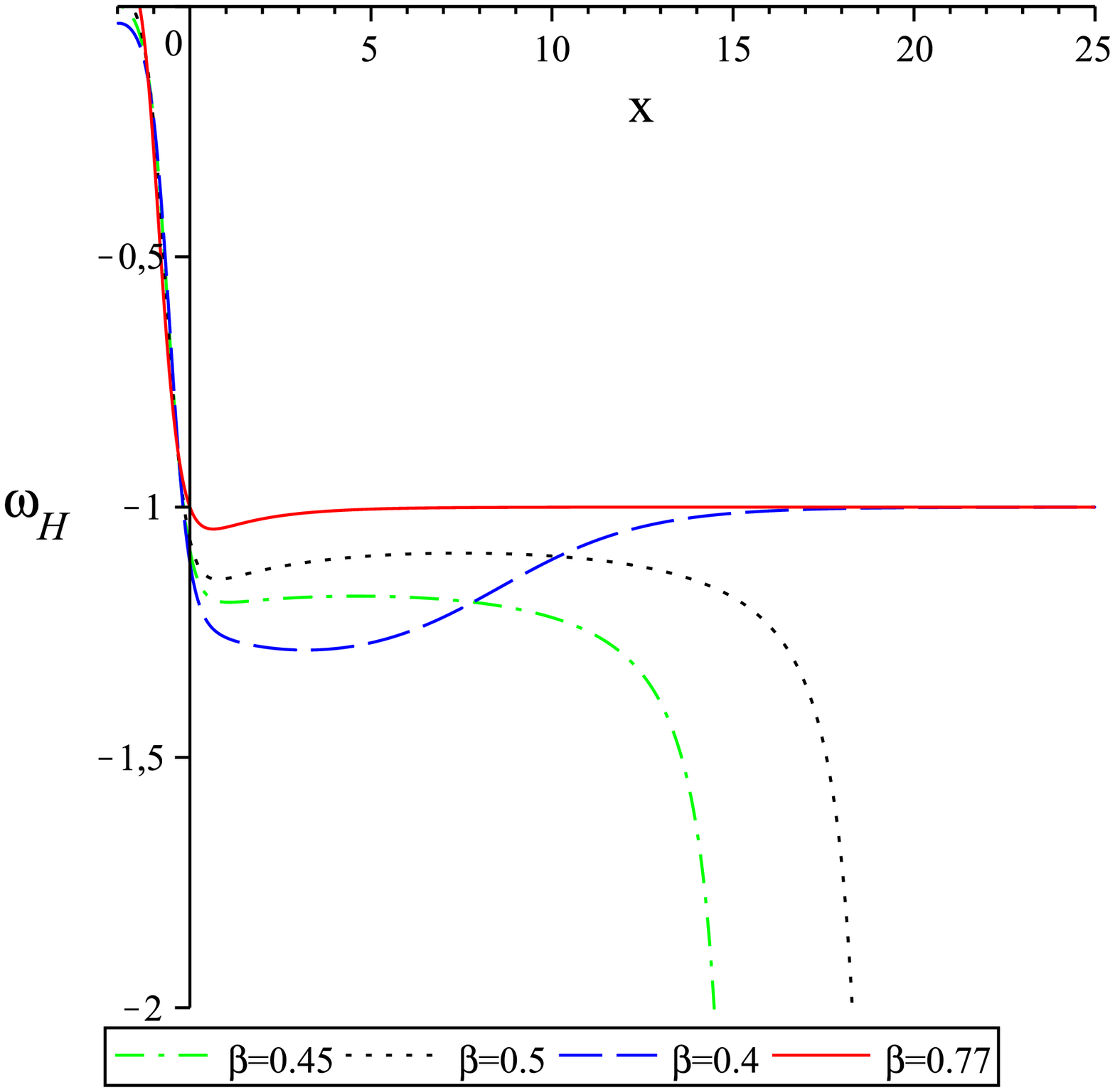}
\end{center}
\caption{ Parameter $\protect\omega _{_{\mathrm{H}}}$ against $x$ for the normal branch ($0.43<\beta$) and for the self-accelerating branch ($\beta<0.43$). The cosmological parameters are assumed to be $\Omega _{m}=0.27$, $q_{0}=-0.7$ and $\Omega_\alpha=0.1$. The plot has been done for several values of the holographic parameter $\beta$ that are stated at the bottom of the figure. The reason we show this plot in terms of $x$, rather than $z$, is to see what are the effects of the big freeze on the equation of state parameter.}
\label{figwGB}
\end{figure}


\subsubsection{Vanishing discriminant ($\mathcal{D}=0$)}

Finally, the discriminant $\mathcal{D}$ vanishes when $\beta=\beta_+$ or $%
\beta_-$. This can take place only on the normal branch. In this fine tuned
case, the solution of the modified Friedmann equation (\ref{FriedasympGB})
can be expressed as
\begin{equation}
E=\frac{1}{4\Omega_\alpha\sqrt{\Omega_{r_c}}}\left[(2\beta-1)-\frac{1}{%
\mathcal{C}_2+\frac{1}{2\beta}(x-x_2)}\right],  \label{solution2 D=0}
\end{equation}
where
\begin{equation}
\mathcal{C}_2=\frac{1}{2\beta-1-4\Omega_\alpha\sqrt{\Omega_{r_c}}E_2},
\end{equation}
and $x_2$ and $E_2$ are integration constants. By taking the limit $%
x\rightarrow\infty$, the dimensionless Hubble rate (\ref{solution2 D=0})
reduces to
\begin{equation}
E_+=\frac{2\beta-1}{4\Omega_\alpha\sqrt{\Omega_{r_c}}},  \label{E_{+3}}
\end{equation}
Not surprisingly, $E_+$ coincides with the one defined in Eq.~(\ref{E_{+2}})
for vanishing $\mathcal{D}$ on the normal branch. The dimensionless Hubble
rate is positive only for values of $\beta$ such that $\beta=\beta_-$ where
the brane is asymptotically de Sitter. The solution (\ref{solution2 D=0}) is
not valid on the normal expanding branch with $\beta=\beta_+$, as the Hubble
rate becomes negative, a condition which signals that the asymptotical
behaviour given by (\ref{solution2 D=0}) breaks down. In fact, what happens
on the normal branch for $\beta=\beta_+$ is that the brane hits a big freeze
singularity at a finite scale factor and the brane expansion can be
approximated by Eq.~(\ref{solution2 D>0}).

\subsection{Numerical analysis}

The asymptotical analysis we have carried out in the previous subsection can
be complemented with a numerical analysis.  We show some examples
of the numerical analysis in Fig {\ref{E with GB}}. More precisely, we
obtain:

\begin{itemize}
\item If $\beta<\beta_{\mathrm{lim}}$, the solution corresponds to the
self-accelerating branch and the brane is asymptotically de Sitter in the
future.

\item If $\beta _{\mathrm{lim}}<\beta <\beta _{-}$, the solution corresponds
to the normal branch and the brane faces a big freeze in the future. Notice
that for the case $\beta =1/2$ with $\Omega_\alpha\neq 0$ the brane faces a big freeze, while it faces a
little rip in the absence of a GB term in the bulk ($\Omega_\alpha= 0$).

\item If $\beta_- \leq \beta$, the solution corresponds to the normal branch
and is asymptotically de Sitter in the future.
\end{itemize}

In summary, we conclude that the big rip that might show up on
the self-accelerating branch without GB effects in the bulk is removed by
the inclusion of GB effects in the bulk. However, the situation in the
normal branch is quite different as the presence of the GB effects, even
though they remove the big rip and little rip, it cannot always prevent the presence of a big freeze
singularity in the future evolution of the brane.

\begin{figure}[t]
\begin{center}
\includegraphics[width=0.7\columnwidth]{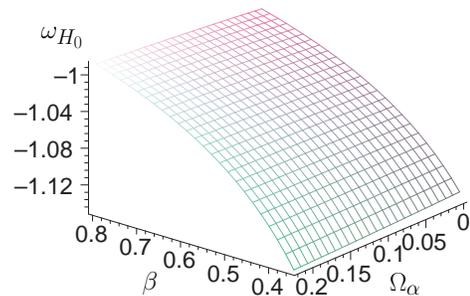}
\end{center}
\caption{Plot of $\protect\omega _{\mathrm{H}_{0}}$ against $\protect\beta $
and $\Omega _{\protect\alpha }$ . The cosmological parameters are assumed to
be $\Omega _{m}=0.27$ and $q_{0}=-0.7$.}
\label{w0}
\end{figure}

For completeness, we will discuss the behaviour of the equation of state $%
\omega _{\mathrm{H}}$ associated with the Ricci HDE. We can show that the
function $\omega _{\mathrm{H}}$ approaches $-1$ at very late-time, in the
case where the brane is asymptotically de Sitter, while it diverges to minus
infinity when the brane face a big freeze singularity in the future (see also Fig.~\ref{figwGB}). In
addition, we can show by using equations (\ref{variation of E}) and (\ref{eq
wH}), that the equation of state parameter $\omega _{\mathrm{H}}$ can be
expressed at the present time as:
\begin{eqnarray}
&\,&\omega _{\mathrm{H}_{0}}=  \notag \\
&\,&\frac{\allowbreak \left[ \beta(1-q_0) +\Omega _{m}\right] \left[
q_{0}(1+3\Omega _{\alpha })-2\right] -(1+q_{0})\allowbreak \left( \Omega
_{\alpha }-1\right) }{3\beta (1-q_{0})(\Omega _{\alpha }+1)}.  \notag \\
\end{eqnarray}%
Fig.~\ref{w0} shows the variation of $\omega _{\mathrm{H}_{0}}$ with
respect to $\Omega _{\alpha }$ and $\beta ,$ where we have used the current
values of $q_{0}\sim -0.7$ and $\Omega _{m}\sim 0.27.$ The figure shows that
$\omega _{\mathrm{H}_{0}}$\ is always less than $-1/3$ at present.  Notice as well that the most likely value of $\beta$ is of the order of $\beta_-$ or larger, and therefore the big freeze singularity will be avoided in the future evolution of the brane (cf. Figs.~\ref{plotbetas} and \ref{w0}).


One can conclude that for all values of $\Omega _{\alpha }$ and $\beta ,$
the equation of state of dark energy $\omega _{\mathrm{H}}$ will evolve at
future in the region $\omega _{\mathrm{H}}<-1/3,$ and converges to $-1$ or
minus infinity as discussed above.  Fig.~{\ref{figwGB}} illustrates
the case of $\ \Omega _{\alpha }=0.1$. This figure shows that the curve of $%
\omega _{\mathrm{H}}$ crosses currently the boundary of the
cosmological constant $\omega _{\mathrm{H}}=-1$ for $\beta _{\mathrm{0}%
}\sim 0.77,$ while for $\beta <\beta _{\mathrm{0}}$ the equation of state of
dark energy will evolve as a phantom-like fluid, i.e. $\omega _{\mathrm{H}}<-1$. This
behaviour will manifest itself more and more in the future until ending with
a big freeze for $\beta _{\mathrm{\lim }}{<}\beta <\beta _{-}$, and with a
de Sitter-like universe for $\beta <\beta _{\mathrm{\lim }}$ and  $\beta_-\leq \beta$.

We note here that according to Eq.~(\ref{boundsbeta}) $\beta$ is always less
than $\beta _{\mathrm{0}}$ and the nature of the Ricci dark energy has always
a phantom-like behaviour.

\begin{figure}[t]
\begin{center}
\includegraphics[width=0.7\columnwidth]{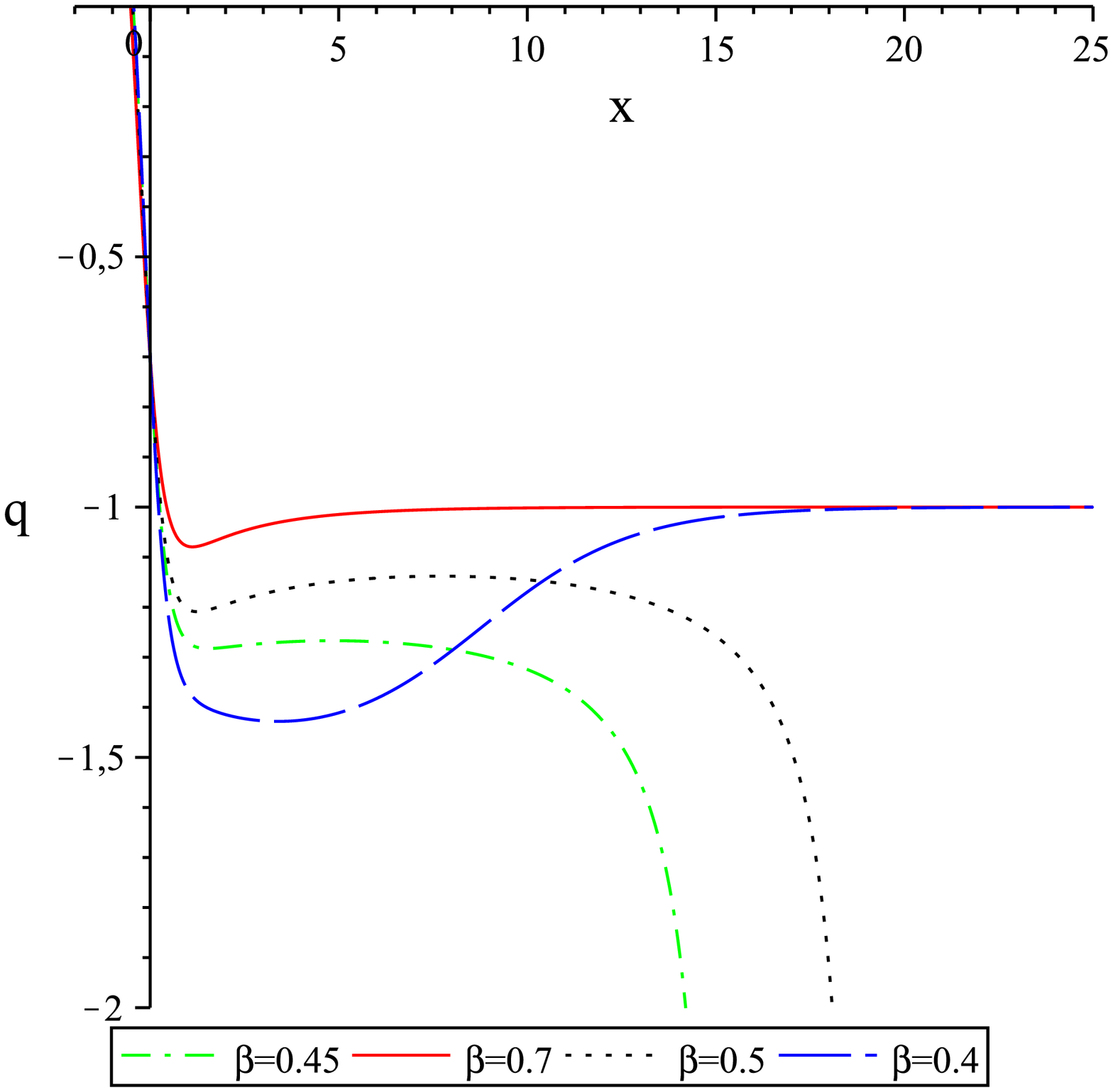}
\end{center}
\caption{ Plot of the deceleration parameter
against $x$ for the normal branch ($0.43<\beta$) and for the self-accelerating branch ($\beta<0.43$). The cosmological parameters are assumed to be $\Omega _{m}=0.27$, $q_{0}=-0.7$ and $\Omega_\alpha=0.1$. The plot has been done for several values of the holographic parameter $\beta$ that are stated at the bottom of the figure. The reason we show this plot in terms of $x$, rather than $z$, is to see what are the effects of the big freeze on the decceleration parameter.}
\label{fig q with GB}
\end{figure}

Once again one can test this model by plotting the deceleration parameter $q$
versus the redshift for some values of $\beta $. Fig.~{\ref{fig q with GB}}
shows that our universe is currently in an accelerating expansion phase.

\section{Conclusions}

In this paper we have considered an holographic DGP brane-world model with
the average radius of the Ricci scalar curvature as an infrared cutoff. We have
shown that the late-time acceleration of the Universe with and without the
Gauss-Bonnet (GB) term in the bulk, is consistent with the current
observational data.

The Ricci HDE is characterised by the parameter $\beta$ (see Eq.~(\ref{HRDE}%
)) which is bounded by the current value of the dimensionless Ricci HDE $%
\Omega _{\mathrm{H}_{0}}$ (cf. Eq.~(\ref{boundsbeta})). In addition, there
is a limiting value of $\beta$, which we named $\beta_{\mathrm{lim}}$ (see
Eqs.~(\ref{conditions for beta}) and (\ref{betalim})) that splits the allowed
$\beta$ values for the self-accelerating branch ($\epsilon=1$) from the
normal branch ($\epsilon=-1$). More precisely, values of $\beta$ such that $%
\beta<\beta_{\mathrm{lim}}$ correspond to the self-accelerating branch while values of
$\beta$ such that $\beta_{\mathrm{lim}}<\beta$ corresponds to the normal
branch. We can estimate that $\beta_{\mathrm{lim}}\sim 0.43$ assuming that
our model does not deviate too much from the $\Lambda$CDM model at present.
The value acquired by $\beta$ is very important in determining the
asymptotical behaviour of the brane and that of the Ricci HDE.

In the absence of GB effects, the equation of state of the Ricci HDE will
evolve in the region $\omega _{\mathrm{H}}<-1$ and its late-time behaviour
is phantom-like with two distinct destinies of the universe: (i) for $%
1/2<\beta $, corresponding exclusively to the normal branch, this
phantom-like behaviour will be more and more like a cosmological constant in
the far future and the brane expansion is asymptotically de Sitter. In this
case, the normal branch does not require the inclusion of a GB term in the
bulk to remove any singularity in the far future. (ii) for $\beta \leq 1/2$
(self-accelerating or normal branch), this phantom-like behaviour continues
and implies a super-accelerated expansion on the brane leading to a big rip
or a little rip singularity. These sets of solutions may require a GB term in
order to remove or smooth those singularities. This is what motivates us to
consider a GB correction in the bulk.

The introduction of a GB term in the bulk may  remove the big rip
and little rip from the future evolution of the brane. In the
self-accelerating branch the brane becomes asymptotically de Sitter,
however, the normal branch can hit a big freeze singularity in the far
future.  More precisely, the normal branch is asymptotically de
Sitter for $\beta _{-}<\beta $ and hits a big freeze singularity when $\beta
_{\mathrm{lim}}\leq \beta <\beta _{-}$ (see Eqs.~(\ref{defbetapm}), (4.7) and (\ref{solution2 D>0})).
The Ricci HDE has a phantom-like behaviour which  either approaches a de Sitter-like state or induces a big freeze as explained above. The case corresponding to a big freeze singularity is quite unlikely because it leads to a quite negative equation of state at present (please see Figs.~\ref{plotbetas} and  {\ref{w0}).  At this regard, we would like to stress that an observationally consistent  value of the equation of state (EOS) parameter of the Ricci HDE is even more important than the GB parameter to reject the parameter region where the future singularity appears. In fact, even in the absence of a GB parameter, the EOS parameter already pick up a singularity free region (cf. Fig.~2). Then what is the role of the GB term? it narrows the region affected by the singularity if $\beta\leq 1/2$ (with respect to the model without a GB term) and enlarges that region if $1/2\leq \beta$ (with respect to the model without a GB term).

In summary, it is always possible to construct a Ricci HDE brane-world model
consistent with current observations. The brane with or without GB effects
is always free of future singularities except for the range of values of $%
\beta$ such that $\beta_{\rm{lim}}\leq \beta \leq 1/2$ which happens always in the
normal branch. However, these values are very unlikely (cf. Figs.~\ref{plotbetas} and  {\ref{w0}) because they lead to a very negative equation of state at the present as explained above.

This study opens a new way of considering interactions between
dark matter and the holographic dark energy. We hope to report this issue in a
forthcoming publication \cite{preparation2}.

\acknowledgments

The authors are grateful to E.-H. El-Boudouti and A. Vale for a careful reading of the
manuscript. M.B.L. is supported by the Portuguese Agency Fund\c{c}\~{a}o
para a Ci\^{e}ncia e Tecnologia through SFRH/BPD/26542/2006 and PTDC/FIS/111032/2009.
A.E. and T.O. are supported by CNRST, through the fellowship URAC 07/214410.


\begin{thebibliography}{99}
\bibitem{Perlmutter:1998np} S.~Perlmutter et al.,
Astrophys.\ J.\ \textbf{517}, 565 (1999) [arXiv:astro-ph/9812133].

A.~G.~Riess et al., 
Astron.\ J.\ \textbf{116}, 1009 (1998) [arXiv:astro-ph/9805201].

M.~Kowalski \textit{et al.},
Astrophys.\ J.\ \textbf{686}, 749 (2008) [arXiv:0804.4142 [astro-ph]].

\bibitem{Komatsu:2010fb} D.~N.~Spergel et al.,
Astrophys.\ J.\ Suppl.\ \textbf{148}, 175 (2003) [arXiv:astro-ph/0302209];
ibid. 
Astrophys.\ J.\ Suppl.\ \textbf{170}, 377 (2007) [arXiv:astro-ph/0603449];
E.~Komatsu \textit{et al.} [WMAP Collaboration],
Astrophys.\ J.\ Suppl.\ \textbf{180}, 330 (2009) [arXiv:0803.0547
[astro-ph]].

\bibitem{Tegmark2004} M. Tegmark, et al., SDSS Collaboration, Phys. Rev. D
\textbf{69}, 103501 (2004)[arXiv:astro-ph/0310723];
J.K. Adelman-McCarthy, et al., SDSS Collaboration, Astrophys. J. Suppl. \textbf{175}, 297 (2008) [arXiv:0707.3413 [astro-ph]].

\bibitem{Peebles} P.J.E. Peebles and B. Ratra, Rev. Mod. Phys. \textbf{75}, 559 (2003) [astro-ph/0207347].

\bibitem{Ratra} B. Ratra and P.J.E. Peebles, Phys. Rev. D \textbf{37}, 3406 (1988); I. Zlatev, L. Wang and P.J. Steinhardt, Phys. Rev. Lett. \textbf{82}, 896 (1999).

\bibitem{ARMENDARIZ-PICON 2001} Amendaiz-Picon,
Phys. Rev. D \textbf{63}, 103510 (2001).

\bibitem{Caldwell} R.R. Caldwell, Phys. Lett. B \textbf{545}, 23 (2002); S.M. Carroll, M. Hoffman and M. Trodden, Phys. Rev. D \textbf{68}, 023509 (2003); J. G. Hao, X. Z. Li, Phys. Rev. D \textbf{68}, 043501 (2003);
D. J. Liu, X. Z. Li, Phys. Rev. D \textbf{68}, 067301 (2003).

\bibitem{'tHooft:1993gx} G.~'t Hooft,
arXiv:gr-qc/9310026.

\bibitem{Susskind:1994vu} L.~Susskind, 
J.\ Math.\ Phys.\ \textbf{36}, 6377 (1995) [arXiv:hep-th/9409089].

\bibitem{Cohen:1998zx} A.~G.~Cohen, D.~B.~Kaplan and A.~E.~Nelson,
Phys.\ Rev.\ Lett.\ \textbf{82}, 4971 (1999) [arXiv:hep-th/9803132].

\bibitem{Li:2004rb} M. Li, Phys. Lett. B \textbf{603}, 1 (2004)
[arXiv:hep-th/0403127].

\bibitem{Hsu:2004ri} S.~D.~H.~Hsu, 
Phys.\ Lett.\ B \textbf{594}, 13 (2004) [arXiv:hep-th/0403052].

\bibitem{Wu:2007tp} X.~Wu, R.~G.~Cai and Z.~H.~Zhu,
Phys.\ Rev.\ D \textbf{77}, 043502 (2008) [arXiv:0712.3604 [astro-ph]].


\bibitem{preparation} M. Bouhmadi-L\'{o}pez, A. Errahmani and T. Ouali,
Phys. Rev. D \textbf{84}, 083508 (2011) [arXiv:1104.1181 [astro-ph.CO]].

\bibitem{saridakis} E.~N.~Saridakis,
  Phys.\ Lett.\ B {\bf 660}, 138 (2008)
  [arXiv:0712.2228 [hep-th]]; ibid JCAP {\bf 0804}, 020 (2008)
  [arXiv:0712.2672 [astro-ph]]; ibid Phys.\ Lett.\ B {\bf 661}, 335 (2008)
  [arXiv:0712.3806 [gr-qc]].

\bibitem{Gao} C. Gao, F.Q. Wu, X. Chen, Y.G. Shen, Phys. Rev. D \textbf{79}, 043511 (2009) [arXiv: 0712.1394 [astro-ph]].

\bibitem{Nojiri:2005pu} 
S.~'i.~Nojiri and S.~D.~Odintsov,
  Gen.\ Rel.\ Grav.\  {\bf 38}, 1285 (2006)
  [hep-th/0506212].

\bibitem{Dvali:2000hr} G.~R.~Dvali, G.~Gabadadze, M.~Porrati,
Phys.\ Lett.\ \textbf{B484}, 112 (2000) [hep-th/0002190]; ibid.
Phys.\ Lett.\ \textbf{B485}, 208 (2000) [hep-th/0005016].

\bibitem{Deffayet:2000uy} C.~Deffayet,
Phys.\ Lett.\ B \textbf{502}, 199 (2001) [arXiv:hep-th/0010186].

\bibitem{BouhmadiLopez:2007ts} V.~Sahni and Y.~Shtanov,
JCAP \textbf{0311}, 014 (2003) [arXiv:astro-ph/0202346]; L.~P.~Chimento,
R.~Lazkoz, R.~Maartens, I.~Quiros,
JCAP \textbf{0609}, 004 (2006) [astro-ph/0605450]; H.~S.~Zhang and
Z.~H.~Zhu,
Phys.\ Rev.\ D \textbf{75}, 023510 (2007) [arXiv:astro-ph/0611834];
M.~Bouhmadi-L\'{o}pez, R.~Lazkoz, 
Phys.\ Lett.\ \textbf{B654}, 51-57 (2007). [arXiv:0706.3896 [astro-ph]].

\bibitem{BouhmadiLopez:2010pp} M.~Bouhmadi-L\'{o}pez,
JCAP \textbf{0911}, 011 (2009) [arXiv:0905.1962 [hep-th]]; M.~Bouhmadi-L\'{o}%
pez, S.~Capozziello, V.~F.~Cardone,
Phys.\ Rev.\ \textbf{D82}, 103526 (2010) [arXiv:1010.1547].

\bibitem{Kofinas:2003rz} G.~Kofinas, R.~Maartens and E.~Papantonopoulos,
JHEP \textbf{0310}, 066 (2003) [arXiv:hep-th/0307138].

\bibitem{richard} R.~A.~Brown, R.~Maartens, E.~Papantonopoulos and
V.~Zamarias,
JCAP \textbf{0511}, 008 (2005) [arXiv:gr-qc/0508116]; R.~A.~Brown,
Gen.\ Rel.\ Grav.\ \textbf{39}, 477 (2007) [arXiv:gr-qc/0602050]; R.~G.~Cai,
H.~S.~Zhang and A.~Wang,
Commun.\ Theor.\ Phys.\ \textbf{44}, 948 (2005) [arXiv:hep-th/0505186].

\bibitem{BouhmadiLopez:2008nf} M.~Bouhmadi-L\'{o}pez and P.~V.~Moniz,
Phys.\ Rev.\ D \textbf{78}, 084019 (2008) [arXiv:0804.4484 [gr-qc]];
M.~Bouhmadi-L\'{o}pez, Y.~Tavakoli and P.~V.~Moniz,
JCAP \textbf{1004}, 016 (2010) [arXiv:0911.1428 [gr-qc]].

\bibitem{Zhang} Chao-Jun Feng \ and Xin Zhang
Phys. Lett. B \textbf{680}, 399 (2009) [arXiv:gr-qc/0904.0045v3].

\bibitem{Shtanov:2002ek} 
  Y.~Shtanov and V.~Sahni,
  Class.\ Quant.\ Grav.\  {\bf 19}, L101 (2002)
  [gr-qc/0204040].


\bibitem{Barrow:2004xh}
  J.~D.~Barrow,
  Class.\ Quant.\ Grav.\  {\bf 21}, L79 (2004)
  [gr-qc/0403084].

\bibitem{Nojiri:2005sx} S.~Nojiri, S.~D.~Odintsov and S.~Tsujikawa,
Phys.\ Rev.\ D \textbf{71}, 063004 (2005) [arXiv:hep-th/0501025].

\bibitem{Cattoen:2005dx} C.~Catto\"{e}n and M.~Visser,
Class.\ Quant.\ Grav.\ \textbf{22}, 4913 (2005) [arXiv:gr-qc/0508045]; L.~Fern%
\'{a}ndez-Jambrina and R.~Lazkoz,
Phys.\ Rev.\ D \textbf{74}, 064030 (2006) [arXiv:gr-qc/0607073].

\bibitem{bigrip} R. R. Caldwell, Phys. Lett. B \textbf{545}, 23 (2002) [arXiv:astro-ph/9908168];
A. A. Starobinsky, Grav. Cosmol. \textbf{6}, 157 (2000) 157 [arXiv:astro-ph/9912054];
S. M. Carroll, M. Hoffman and M. Trodden, Phys. Rev. D \textbf{68}, 023509  (2003) [arXiv:astro-ph/0301273];
R. R. Caldwell, M. Kamionkowski and N. N. Weinberg, Phys. Rev. Lett. \textbf{91}, 071301  (2003) [arXiv:astro-ph/0302506];
 M.~Bouhmadi-L\'opez and J.~A.~Jim\'enez Madrid,
  JCAP {\bf 0505}, 005 (2005)
  [astro-ph/0404540];
E.~Elizalde, S.~'i.~Nojiri and S.~D.~Odintsov,
  Phys.\ Rev.\ D {\bf 70}, 043539 (2004)
  [hep-th/0405034].


\bibitem{BouhmadiLopez:2006fu} M.~Bouhmadi-L\'opez, P.~F.~Gonz\'alez-D\'{\i}%
az, P.~Mart\'{\i}n-Moruno, 
Phys.\ Lett.\ \textbf{B659}, 1 (2008) [gr-qc/0612135]; ibid. Int.\ J.\
Mod.\ Phys.\ \textbf{D17}, 2269 (2008) [arXiv:0707.2390 [gr-qc]].

\bibitem{Stefancic:2004kb} H.~\v{S}tefan\v{c}i\'c,
Phys.\ Rev.\ \textbf{D71}, 084024 (2005) [astro-ph/0411630].

\bibitem{BouhmadiLopez:2005gk} M.~Bouhmadi-L\'opez,
Nucl.\ Phys.\ \textbf{B797}, 78 (2008) [astro-ph/0512124].

\bibitem{Frampton:2011sp} P.~H.~Frampton, K.~J.~Ludwick, R.~J.~Scherrer,
Phys.\ Rev.\ \textbf{D84}, 063003 (2011) [arXiv:1106.4996 [astro-ph.CO]]; 
I.~Brevik, E.~Elizalde, S.~Nojiri and S.~D.~Odintsov,
  Phys.\ Rev.\ D {\bf 84}, 103508 (2011)
  [arXiv:1107.4642 [hep-th]];  P.~H.~Frampton, K.~J.~Ludwick, S.~'i.~Nojiri, S.~D.~Odintsov and R.~J.~Scherrer,
  arXiv:1108.0067 [hep-th]; S.~'i.~Nojiri, S.~D.~Odintsov and D.~S\'aez-G\'omez,
  arXiv:1108.0767 [hep-th].

\bibitem{preparation2} M.-H. Belkacemi, \ A. Errahmani, and T. Ouali, in
preparation.
\end{thebibliography}
\end{document}